\documentclass[ pre,twocolumn,floatfix,amsmath,amssymb,aps,showpacs,longbibliography,superscriptaddress]{revtex4-1}
\usepackage{graphicx}
\usepackage{xcolor}
\usepackage{siunitx}
\usepackage{hyperref}
\usepackage[normalem]{ulem}
\usepackage{here}
\usepackage {amsmath}
\usepackage{mathrsfs}
\usepackage{bm}
\usepackage{color} \usepackage{graphicx} \graphicspath{{Figures/}}
\usepackage{comment} \usepackage{color}
\usepackage[normalem]{ulem}
\renewcommand{\epsilon}{\varepsilon}

\newcommand{\tn}[1]{\textnormal{#1}} 

\begin{document}

\title {Interfacial bacteria-driven rotation of free-floating active droplets}

\author{Dixi Yang}
\thanks{These authors contributed equally}
\affiliation{Advanced Materials Thrust, The Hong Kong University of Science and Technology (Guangzhou), Guangzhou 511453, China}

\author{Anheng Wang}
\thanks{These authors contributed equally}
\affiliation{Institute of Chinese Medical Sciences \& State Key Laboratory of Mechanism
and Quality of Chinese Medicine, University of Macau, Macau SAR 999078, China}
\affiliation{Department of Pharmacy and Pharmaceutical Sciences, Faculty of Medicine, University of Macau, Macau SAR 999078, China}
\affiliation{Zhuhai UM Science and Technology Research Institute, University of Macau, Hengqin 519031, Guangdong, China}
\affiliation{University of Macau Advanced Research Institute in Hengqin, Hengqin 519031, Guangdong, China}

\author{Chunming Wang}
\email{cmwang@um.edu.mo}
\affiliation{Institute of Chinese Medical Sciences \& State Key Laboratory of Mechanism
and Quality of Chinese Medicine, University of Macau, Macau SAR 999078, China}
\affiliation{Department of Pharmacy and Pharmaceutical Sciences, Faculty of Medicine, University of Macau, Macau SAR 999078, China}
\affiliation{Zhuhai UM Science and Technology Research Institute, University of Macau, Hengqin 519031, Guangdong, China}
\affiliation{University of Macau Advanced Research Institute in Hengqin, Hengqin 519031, Guangdong, China}

\author{Hajime Tanaka}
\email{tanaka@iis.u-tokyo.ac.jp}
\affiliation{Research Center for Advanced Science and Technology, The University of Tokyo, 4-6-1 Komaba, Meguro-ku, Tokyo 153-8904, Japan}
\affiliation{Department of Fundamental Engineering, Institute of Industrial Science, The University of Tokyo, 4-6-1 Komaba, Meguro-ku, Tokyo 153-8505, Japan}

\author{Jiaxing Yuan}
\email{jiaxingyuan@hkust-gz.edu.cn}
\affiliation{Advanced Materials Thrust, The Hong Kong University of Science and Technology (Guangzhou), Guangzhou 511453, China}
\affiliation{The Hong Kong University of Science and Technology, Clear Water Bay, Hong Kong SAR, China}

\date{October 18, 2026}

\begin{abstract}
Active droplets provide a platform for converting microscopic propulsion into macroscopic flow. Previous realizations of bacteria-driven droplet rotation typically rely on nearby solid boundaries, external fields, or dense active nematics. Here, we report autonomous rotation of droplets in a three-dimensional aqueous two-phase system driven by interfacial population of \textit{Pseudomonas aeruginosa}. Combining confocal microscopy with fluid particle dynamics simulations, we show that amphiphilic bacteria accumulate at the dextran--polyethylene glycol interface and preferentially move and orient along the local tangent. Interfacial localization partitions the force-free swimming dipole between the two liquid phases, allowing either phase to experience a finite active torque while the combined system remains torque balanced. The simulated interfacial population is more spatially heterogeneous than in the experiments, emphasizing that both the bacterial distribution and its signed orientational organization can influence the instantaneous torque. Over the accessible size range, the macroscopic angular velocity exhibits an inverse-radius ($R^{-1}$) dependence. We also formulate a theoretical model to rationalize this scaling law, revealing how the macroscopic rotation is governed by collective interfacial swimming and surface coverage. These results identify a fluid-interface-mediated route for converting swimmer activity into macroscopic rotational flow and establish a strategy for harnessing interfacial active matter to regulate droplet dynamics.
\end{abstract}

\maketitle

\section{Introduction}
\label{sec:introduction}

Active matter comprises units that locally consume energy to generate motion or mechanical stresses, thereby sustaining collective dynamics far from thermodynamic equilibrium~\cite{marchetti2013hydrodynamics,bechinger2016active,liu2020oscillating,shaebani2020computational,grober2023unconventional}. Coupling such activity to a deformable fluid interface can generate motion on the scale of an entire droplet, making active droplets powerful model systems for investigating how microscopic forcing is converted into organized fluid flow~\cite{Singh2019,arnold2023active}. Several mechanisms mediate this conversion: chemically active droplets self-propel via solute transport coupled with Marangoni flows~\cite{michelin2023self}, whereas droplets containing active nematics translate and rotate through extensile stresses and orientational-defect dynamics~\cite{gao2017self}. Furthermore, chemically active nematic liquid-crystal droplets exhibit spontaneous rotation driven by the coupling between nematic order and surfactant transport~\cite{morozov2019orientational}.

Motile bacteria offer an experimentally accessible and versatile source of microscopic activity. Asymmetric geometries and ratchets can rectify bacterial swimming into persistent rotation \cite{sokolov2010swimming,dileonardo2010bacterial}, while depletion interactions can assemble bacteria into rotating clusters whose angular velocity scales inversely with cluster size \cite{schwarz2012phase}. When confined within fluid droplets, the conversion of bacterial motility into macroscopic motion depends heavily on collective dynamics and environmental constraints. Magnetotactic bacteria, for instance, form functional rotary motors under an applied magnetic field and transmit torque to the surrounding oil phase \cite{vincenti2019magnetotactic}. Similarly, dense suspensions of \textit{Escherichia coli} can drive droplet motion consistent with slippery rolling over nearby substrates \cite{ramos2020bacteria}, while external liquid-crystal environments can direct the locomotion of bacteria-laden droplets \cite{rajabi2021directional}. Fluid interfaces can also host active bacterial populations that directly alter droplet dynamics; notably, Dewangan \textit{et al.} showed that bacteria adhering to oil-droplet surfaces induce droplet rotation by relying on hydrodynamic interactions (HIs) with a nearby solid boundary to break rotational symmetry \cite{dewangan2019rotating}. These systems illustrate how confinement, boundary asymmetry, and bacterial activity can be harnessed to drive droplet-scale rotation.

The behavior of bacteria at a fluid interface extends far beyond propulsion. For instance, interfacial bacterial films can undergo rheological transitions, evolving from active fluids into viscoelastic layers~\cite{vaccari2015films,vaccari2017films}. Furthermore, distinct strains of \textit{Pseudomonas aeruginosa} exhibit qualitatively different phenotypic responses at identical oil--water boundaries, forming either motile populations or elastic films~\cite{niepa2017films}. Single-cell tracking reveals that such bacteria can become hydrodynamically trapped or adsorbed at liquid interfaces, displaying swimming trajectories that deviate significantly from those in the bulk~\cite{deng2020motile}. Consequently, both the localization and the active states of bacteria are critical determinants of the stresses they transmit to neighboring fluids.

To expand beyond oil-based emulsions, aqueous two-phase systems (ATPSs)—formed via liquid--liquid phase separation (LLPS) in polymer mixtures—provide a versatile and biocompatible platform for investigating the coupling between phase separation,  activity, and droplet dynamics. LLPS is fundamentally important across a wide range of scientific and engineering disciplines, including soft and active matter \cite{onuki2002phase,glotzer1993physical,liu2021evolution,schmid2022understanding,yuan2023mechanical,yuan2024colloid,yuan2025network,chen2026entropic}, biological systems \cite{iborra2007can,berry2018physical,Zhang2021,tanaka2022Viscoelastic,yang2025active,chen2026phase}, and industrial applications \cite{Ebbens2011,wang2019progress,Bok2022}. For example, in dextran (DEX) and polyethylene glycol (PEG) ATPS, \textit{Bacillus subtilis} wets the DEX, where their spatial partitioning reflects a balance between bacterial propulsion and interfacial confinement \cite{cheon2024partitioning}.

Previous studies of bacteria-driven droplet rotation have relied on nearby solid boundaries to select the rotational state~\cite{dewangan2019rotating}, dense bacterial suspensions to generate collective active stresses~\cite{ramos2020bacteria,rajabi2021directional}, or external driving fields~\cite{vincenti2019magnetotactic}. A force-free swimmer in a homogeneous bulk fluid exerts no net force or torque on the fluid as a whole. A liquid--liquid interface, however, provides a qualitatively different mechanical setting: when an interfacially trapped swimmer straddles the two phases, the two members of its active force dipole can act predominantly on different phases. The total torque of the combined system remains balanced, while either phase separately can experience a finite torque. The resulting droplet-scale torque depends not only on interfacial localization but also on how the swimmers are distributed over the interface and on whether their tangential motion possesses a nonzero signed circulation. 

Using the same experimental platform as our companion study~\cite{yang2026a}, we focus on this  microscopic mechanical question. We combine three-dimensional confocal imaging with hydrodynamic simulations and theories to connect interfacial localization and tangential motion to the torque exerted separately on the DEX- and PEG-rich phases. The key novelty is not the occurrence of droplet rotation which has been reported in other active systems~\cite{dewangan2019rotating,schwarz2012phase}, but the conversion of a force-free swimmer dipole into phase-resolved torque through interfacial partitioning, together with the possibility that collective interfacial organization controls the resulting macroscopic rotation.

\section{Experimental Methods}
\label{sec:exp_method}

Sample culturing, ATPS preparation, and microscopy follow the procedures established in our companion study~\cite{yang2026a}. The sections below summarize these procedures for completeness. We detail the analysis workflows specific to the rotational dynamics reported here in \cite{Supple}.

\subsection{Bacterial culture}
The bacterial strains used in this study were \textit{Pseudomonas aeruginosa} (\textit{P. aeruginosa}; ATCC 10145) and its green fluorescent protein (GFP)-tagged variant (10145GFP ATCC). Bacterial stocks were stored at $-80^\circ\text{C}$ in Luria-Bertani (LB) broth (Solarbio, L8290). For routine experiments, bacteria were streaked from the frozen stock onto LB agar (Solarbio, L8291) and incubated at $37^\circ\text{C}$ for 24 hours. For liquid cultures, a single colony was inoculated into 10 mL of LB (Solarbio, L8290) and grown overnight at $37^\circ\text{C}$ with vigorous shaking (150 rpm). This overnight starter culture was then subcultured into fresh medium for experimental use. The culture was diluted into fresh LB medium and grown to the desired optical density at 600 nm (OD$_{600}=0.5$--$0.6$, mid-exponential phase). All culture media and reagents were sterilized by autoclaving prior to use. Bacterial biomass was quantified using the packed cell volume (PCV) method. Aliquots of 500~$\mu$L of the bacterial culture, taken at specified time points, were transferred into calibrated 1.5 mL conical centrifuge tubes (Axygen, MCT-150-C-S). The samples were pelleted by centrifugation at 5,000 x g for 15 minutes at room temperature. Following centrifugation, the supernatant was carefully decanted, and the volume of the resulting cell pellet was directly read from the volumetric graduations on the 1.5 mL centrifuge tube. The PCV was expressed as the percentage of the pellet volume relative to the total initial culture volume ($V_{\text{pellet}} / V_{\text{total}} \times 100$). All measurements were performed in triplicate to ensure reproducibility.

\subsection{Preparation and inoculation of aqueous two-phase system (ATPS)}

\textbf{Media preparation.} Stock solutions of Dextran (DEX; MW 450-650 kDa; Sigma-Aldrich, 31392-50G) and Polyethylene glycol (PEG; MW 100 kDa; Sigma-Aldrich, U07G051-100G) were prepared by dissolving them in standard Luria-Bertani broth (LB; Solarbio, L8290) to final concentrations of 5~wt\%. The solutions were sterilized by autoclaving and stored at $4^\circ\text{C}$.

\textbf{Bacterial strain adaptation.} Prior to inoculation into the ATPS, \textit{P. aeruginosa} (ATCC 10145) was pre-adapted to the polymer-rich environment. A single colony was first grown overnight in standard LB at $37^\circ\text{C}$ with shaking (150 rpm). This culture was pelleted by centrifugation (5,000~$\times$~g, 10 min), and the supernatant was discarded. The bacteria were then resuspended in the sterile DEX-LB (DEX-LB) solution (5~wt\%) and incubated overnight at $37^\circ\text{C}$ with shaking (150 rpm) to allow for metabolic adaptation.

\textbf{Biomass quantification and inoculation.} The adapted overnight culture was centrifuged (5,000~$\times$~g, 15 min, $4^\circ\text{C}$). The supernatant was discarded, and the total bacterial biomass was quantified using the PCV method, as previously described.

\textbf{ATPS assembly.} The final ATPS was assembled in sterile 1 mL PCR tubes (Vazyme, PCR00802). The pre-adapted \textit{P. aeruginosa} (ATCC 10145) bacterial pellet was added to the 5~wt\% DEX-LB stock solution, followed by the addition of 5~wt\% PEG-LB stock solution. The components were mixed by pipetting using yellow tips (200~$\mu$L) to achieve the desired final volume fractions for the bacteria ($\phi_{\text{bac}}$), the DEX phase ($\phi_{\text{dex}}$), and the PEG phase ($\phi_{\text{peg}}$). The total volume of the system was 100~$\mu$L.

\subsection{Bright field observation, confocal microscopy, and 3D image acquisition}

Following phase separation, the ATPS containing \textit{P. aeruginosa} (10145GFP ATCC) was gently transferred to confocal dishes (35 mm glass-bottom confocal dish, Nest, 801001). To minimize sample evaporation and preserve the biphasic system, the culture dish was placed in a humidified chamber maintained at $>90$\% relative humidity during time-lapse imaging. Bright field imaging and videography were performed using an Olympus Spin SR equipped with a spinning-disk confocal unit (Yokogawa CSU-W1). Three-dimensional (3D) fluorescence stacks of the GFP-tagged bacteria were acquired using a 100$\times$/1.4 NA oil-immersion objective lens. For 3D reconstructions of the bacterial distribution, $z$-stacks were captured with a step size of 0.5~$\mu$m per slice. For time-resolved measurements, four-dimensional ($xyzt$) time-lapse sequences were acquired in the FITC channel with the same $z$-step of 0.5~$\mu$m.

\section{Simulation Methods}
\label{sec:sim_method}

We propose a particle-field hybrid model for simulating the motion of active bacteria immersed in a binary fluid mixture undergoing demixing, taking into account phase separation dynamics, hydrodynamic interactions (HIs), bacterial activity, and wetting effects. 
Additional simulation analysis methods are provided in \cite{Supple}.

{\bf Bacteria modeling.}
We employed a three-particle model to represent a single bacterium~\cite{furukawa2014activity}. A schematic illustration of the model is in Fig.~S1 in \cite{Supple}. Each bacterium~$i$ consists of a head particle (position~$\bm r_i^{\mathrm{h}}$), a tail particle (position~$\bm r_i^{\mathrm{t}}$), and a virtual phantom particle (position~$\bm r_i^{\mathrm{vir}}$). The head and tail particles are connected by a harmonic spring with equilibrium distance~$\ell_1$ to maintain structural integrity.
The virtual particle's position is calculated using the tail-to-head orientation vector $\hat{\mathbf{D}}=(\bm r_i^{\mathrm{h}}-\bm r_i^{\mathrm{t}})/|\bm r_i^{\mathrm{h}}-\bm r_i^{\mathrm{t}}|$ as $\bm r_i^{\mathrm{vir}} = \bm r_i^{\mathrm{t}} - \ell_2\hat{\mathbf{D}}$. To model active propulsion, an active force of strength~$f_{\text{act}}$ is applied to the tail particle directed toward the head. An equal and opposite counterforce acts on the virtual phantom particle, leading to dipolar flow field around a bacterium~\cite {drescher2011fluid} and ensuring momentum conservation.

{\bf Fluid particle dynamics.} 
Accurately simulating bacterial systems requires explicit consideration of HIs. We incorporate HIs by utilizing the fluid particle dynamics (FPD) method~\cite{tanaka2000,Furukawa2018,Yuan2022} that directly solves the Navier--Stokes (NS) equation. Within the framework of the FPD method, each constituent particle of the bacterium is modeled as a viscous fluid particle with a smooth interfacial profile,
\begin{equation}
\phi_{i}({\bm r})=\frac{1}{2}\left\{\tanh
\left[\frac{a_i-\left|{\bm r}-{\bm r}_{i}\right|}{\xi}\right] + 1\right\},
\label{eq:phi}
\end{equation}
where $a_i$ is the particle radius, $\xi$ the interfacial thickness, and $\bm{r}_i$ the position of the $i$-th particle (including head and tail particles).  
The viscosity field is expressed as
\begin{equation}
\eta(\bm{r}) = \eta_{\rm s} + \left(\eta_{\rm p}-\eta_{\rm s}\right)\sum_{i=1}^{2N} \phi_i(\bm{r}),
\label{eq:eta}
\end{equation}
where $\eta_{\rm s}$ and $\eta_{\rm p}$ are the solvent and particle viscosities, respectively, and $N$ is the total number of bacteria. By choosing $\eta_{\mathrm{p}}=50\eta_{\mathrm{s}}$, we effectively suppress the flow within the particles, allowing us to approximate each particle as a rigid body~\cite{tanaka2000,Furukawa2018}. 
Note that the summation extends to $2N$ because each bacterium is represented by both a head and a tail particle.

\paragraph*{Total free energy.}
The total free energy of the binary fluid containing bacteria is given by~\cite{araki2006wetting,Furukawa2018}
\begin{equation}
\mathscr{F}\{\psi, {\bm{r}_i}\}= \mathscr{F}_b\{\psi\} + \mathscr{F}_s\{\psi, {\bm{r}_i}\} + \mathscr{F}_a\{\psi, {\bm{r}_i}\} + U\{{\bm{r}_i}\},
\label{eq:F}
\end{equation}
where $\psi(\bm{r})$ is the composition order parameter field and $\bm{r}_i$ is the position of particle~$i$.

\paragraph*{Bulk free energy.}
The bulk Ginzburg--Landau (GL) free energy is written as
\begin{equation}
\mathscr{F}_b\{\psi\} = \int d\bm{r} \left[  \frac{1}{2}r_0 \epsilon_0 \psi^2 + \frac{1}{4}u_0 \psi^4 + \frac{k_0}{2}|\nabla \psi|^2 \right],
\label{eq:Fb}
\end{equation}
where $\epsilon_0 = (T-T_{\rm c})/T_{\rm c}$ is the reduced temperature.  
For $r_0\epsilon_0 < 0$ and $u_0>0$, the mixture phase-separates into two distinct phases $A$ and $B$, characterized by 
\begin{equation}
\psi_{\rm equi}^{A,B} = \pm\sqrt{-\frac{r_0\epsilon_0}{u_0}}.
\end{equation}

\paragraph*{Surface wetting and amphiphilic nature.}
The wetting interaction between the $i$-th particle and the surrounding binary fluid is represented by the surface energy contribution $\mathscr{F}_s\{\psi,\mathbf{r}_i\}$~\cite{araki2006wetting}:
\begin{equation}
\mathscr{F}_s\{\psi,\mathbf{r}_i\} = \sum_{i=1}^{2N} \int d\mathbf{r}\,|\nabla\phi_i|\, f_s(\psi,T),
\quad f_s(\psi,T)=\gamma_i\psi.
\end{equation}
Here, $\gamma_i$ corresponds to the surface coupling strength for the head ($\gamma_i=\gamma_1$) or tail ($\gamma_i=\gamma_2$) particle. The sign of $\gamma_i$ dictates the preferential wetting. A positive $\gamma_i$ favors wetting by Phase B, while a negative $\gamma_i$ favors wetting by Phase A. To model the amphiphilic nature of \emph{Pseudomonas aeruginosa} (\textit{P. aeruginosa}), we set the head coupling strength to $\gamma_1 = -4$ and the tail coupling strength to $\gamma_2 = 2$. These opposite couplings should be regarded as a minimal coarse-grained representation that stabilizes an interfacial, tilted swimmer configuration; they are not intended as a literal measurement of distinct head and tail surface chemistries. The mechanism tested below is therefore the partitioning of the active force dipole across the two phases, rather than the specific numerical choice of $\gamma_1$ and $\gamma_2$.

\paragraph*{Particle impermeability.}
To prevent the composition field from penetrating inside the bacterial bodies, we introduce
\begin{equation}
\mathscr{F}_a\{\psi,\bm{r}_i\} = \sum_{i=1}^{2N} \int d\bm{r}\,\chi \phi_i \psi^2,
\label{eq:Fa}
\end{equation}
where $\chi>0$ effectively enforces the condition $\psi \approx 0$ within the volume occupied by the head and tail particles.

\paragraph*{Direct particle interactions.}
Instead of using a dynamical rule, which switches off bacterial activity when bacteria are closer than a cutoff distance, we introduce repulsive interactions involving the virtual particle to explicitly prevent overlaps~\cite{furukawa2014activity}. Therefore, the virtual particle is no longer a strictly ``phantom" site but participates in direct physical interactions. This choice is more realistic because it allows the active propulsion to remain engaged while the increasing repulsive force on the virtual particle provides a smooth and physically sound mechanism for preventing flagella overlap in crowded conditions. The direct particle interactions include the excluded-volume interactions between all the particles (head, tail, and virtual particles) and the harmonic potential between head and tail particles:
\begin{equation}
U\{\bm{r}_i\}=\sum_{i=1}^{3N}\sum_{j\ne i}U_{\rm rep}(r_{ij}) + \sum_\alpha^N U_{\rm bond}\left(\left|\boldsymbol{r}_\alpha^{\mathrm{h}}-\boldsymbol{r}_\alpha^{\mathrm{t}}\right|\right)
\end{equation}
\begin{equation}
U_{\rm rep}(r_{ij})=\epsilon_1\left(\frac{2a_{ij}}{r_{ij}}\right)^{24}
\label{eq:WCA}
\end{equation}
\begin{equation}
U_{\rm bond}(r_{ij})=\frac{1}{2} \epsilon_2\left(\frac{r_{ij}}{\ell_1}-1\right)^2
\end{equation}
where $r_{ij}=|\bm{r}_i-\bm{r}_j|$ is the center-to-center distance, $a_{ij}=(a_i+a_j)/2$ is the effective radius between particles $i$ and $j$, $a_i$ is the radius of particle~$i$, $\epsilon_1$ is the repulsion strength, and $\epsilon_2$ represents the spring constant. 
We note that the upper limit $3N$ reflects the inclusion of head, tail, and virtual particles for $N$ bacteria.

\paragraph*{Force transformation.}
Since the virtual particle's position is not an independent degree of freedom, the direct interaction force calculated at the virtual site $\bm f_i^{\mathrm{vir}} = -{\partial {U}}/{\partial \bm r_i^{\mathrm{vir}}}$ must be redistributed onto the head and tail particles of the bacteria~$i$. This is done by using the chain rule:
\begin{equation}
\bm f_i^{(\mathrm{h,vir})} = -\frac{\partial {U}}{\partial \bm r_i^{\mathrm{vir}}}
\frac{\partial \bm r_i^{\mathrm{vir}}}{\partial \bm r_i^{\mathrm{h}}}
\end{equation}
and
\begin{equation}
\bm f_i^{(\mathrm{t,vir})}
= -\frac{\partial {U}}{\partial \bm r_i^{\mathrm{vir}}}
\frac{\partial \bm r_i^{\mathrm{vir}}}{\partial \bm r_i^{\mathrm{t}}}\;,
\end{equation}
where
\begin{equation}
\frac{\partial \bm r_i^{\mathrm{vir}}}{\partial \bm r_i^{\mathrm{h}}}
= - \frac{\ell_2}{|\bm r_i^{\mathrm{h}} - \bm r_i^{\mathrm{t}}|}
(\bm I - \hat{\bm D}_i \hat{\bm D}_i),
\end{equation}
\begin{equation}
\frac{\partial \bm r_i^{\mathrm{vir}}}{\partial \bm r_i^{\mathrm{t}}}
= \bm I +  \frac{\ell_2}{|\bm r_i^{\mathrm{h}} - \bm r_i^{\mathrm{t}}|}
(\bm I - \hat{\bm D}_i \hat{\bm D}_i),
\end{equation}
and $\bm I$ is the unit tensor. Thus, the redistributed forces on the head and tail are given by
\begin{equation}
\bm f_i^{(\mathrm{h,vir})}
= - \frac{\ell_2}{|\bm r_i^{\mathrm{h}} - \bm r_i^{\mathrm{t}}|}
(\bm I - \hat{\bm D}_i \hat{\bm D}_i)\,\bm f_i^{\mathrm{vir}},
\end{equation}
\begin{equation}
\bm f_i^{(\mathrm{t,vir})}
= \bm f_i^{\mathrm{vir}} + \frac{\ell_2}{|\bm r_i^{\mathrm{h}} - \bm r_i^{\mathrm{t}}|}
(\bm I - \hat{\bm D}_i \hat{\bm D}_i)\,\bm f_i^{\mathrm{vir}}.
\end{equation}
This transformation satisfies $\bm{f}_i^{(\mathrm{h,vir})} + \bm{f}_i^{(\mathrm{t,vir})} = \bm{f}_i^{\mathrm{vir}}$, which indicates that the force~$\bm{f}_i^{\mathrm{vir}}$ is redistributed to the head and tail particles with no net force introduced.

\paragraph*{Hydrodynamics.}
The velocity field $\bm{v}$ satisfies the incompressible NS equation,
\begin{equation}
\rho\left(\frac{\partial}{\partial t} + \bm{v}\cdot\nabla\right)\bm{v}
= \bm{f}_{\rm rev} + \nabla\cdot\bm{\sigma},
\label{eq:NS}
\end{equation}
where $\rho$ is the density and
\begin{equation}
\bm{\sigma} = \eta(\bm{r})\left[\nabla\bm{v}+(\nabla\bm{v})^\mathsf{T}\right] - p\bm{I}
\end{equation}
is the viscous stress tensor.  $\bm{I}$ is the unit tensor and the pressure $p$ is determined to satisfy the incompressible condition $\nabla\cdot\bm{v}=0$.

The reversible force density is
\begin{equation}
\bm{f}_{\rm rev}(\bm{r}) = -\psi\nabla\mu(\bm{r}) - \sum_{i=1}^{2N} \frac{\phi_i}{\Omega_i}\bm{f}_i,
\label{eq:f_rev}
\end{equation}
ensuring energy dissipation $d\mathscr{F}/dt \le 0$ while conserving momentum~\cite{Furukawa2018}.  
Here $\mu=\delta\mathscr{F}/\delta\psi$ is the chemical potential, $\bm{f}_i=\partial\mathscr{F}/\partial\bm{r}_i$ the force on particle $i$, and $\Omega_i=\int d\bm{r}\,\phi_i$ its effective volume.
Equation~\eqref{eq:NS} is solved using the Marker-and-Cell (MAC) method~\cite{Harlow1965} on a staggered grid with periodic boundaries.  

\paragraph*{Composition dynamics.}
The order parameter evolves according to
\begin{equation}
\frac{\partial\psi}{\partial t} = -\nabla\cdot\bm{j}_\psi,
\end{equation}
with the flux
\begin{equation}
\bm{j}_\psi = \psi\bm{v} - L_\psi\nabla\mu,
\quad L_\psi = L_0\Bigl(1-\sum_i^N \phi_i\Bigr),
\label{eq:flux}
\end{equation}
where $L_0$ is a constant mobility coefficient and $L_\psi$ vanishes inside the particles.

\paragraph*{Time integration.}
Particle positions are updated as
\begin{equation}
\frac{d\bm{r}_i}{dt} = \bm{v}_i,
\quad \bm{v}_i = \frac{\int d\bm{r}\,\bm{v}\phi_i}{\Omega_i}.
\label{eq:ri}
\end{equation}

{\bf Simulation parameters.}
The lengths are measured in units of the lattice spacing $\ell_0$, and time in $\tau_{s}=\rho\ell_0^{2}/\eta_{\rm s}$. We set $a_\text{h}=a_\text{t}=3.2\ell_0$ for head and tail particle, $a_\text{vir}=2.4\ell_0$ for virtual particle, $\xi=\ell_0$, $\eta_{\rm p}=50\eta_{\rm s}$, $\varepsilon_1=10$, $\varepsilon_2=3.2\times10^5$, simulation box size $L=256\ell_0$, $N=400$, $\ell_1=8.0\ell_0$, $\ell_2=9.6\ell_0$, and $\Delta t=2.5\times10^{-3}\tau_{s}$. The GL free-energy parameters are $r_0\epsilon_0=-1$, $u_0=1$, and $k_0=1$.  
The surface coupling strengths are set as $\gamma_1=-4$ and $\gamma_2=2$ for the head and tail, respectively. The setup yields a volume fraction of bacteria $\phi_{\text{bac}} \approx 1\%$ calculated from the two explicit real particles (head and tail). Since this estimate ignores the volume of the cylindrical bacterial body and the virtual particle (which mimics the flagella), it serves as a conservative lower bound. Accounting for these factors suggests a volume fraction of $\phi_{\text{bac}} \approx 3\%$. The average composition $\bar{\psi}$ is $\bar{\psi}=-0.6$ such that the minority-phase fraction $\psi_A=(1+\bar{\psi})/2$ is $\psi_A=0.2$. Simulations with $f_{\text{act}}=1$ correspond to active bacteria, whereas $f_{\text{act}}=0$ represents non-motile bacteria for the control simulation. All simulations were initialized from a homogeneous state with randomly distributed bacteria. The simulations were performed using NVIDIA A800 GPU.

\section{Results and Discussion}

\begin{figure*}[t!]
\centering \includegraphics[width=15cm]{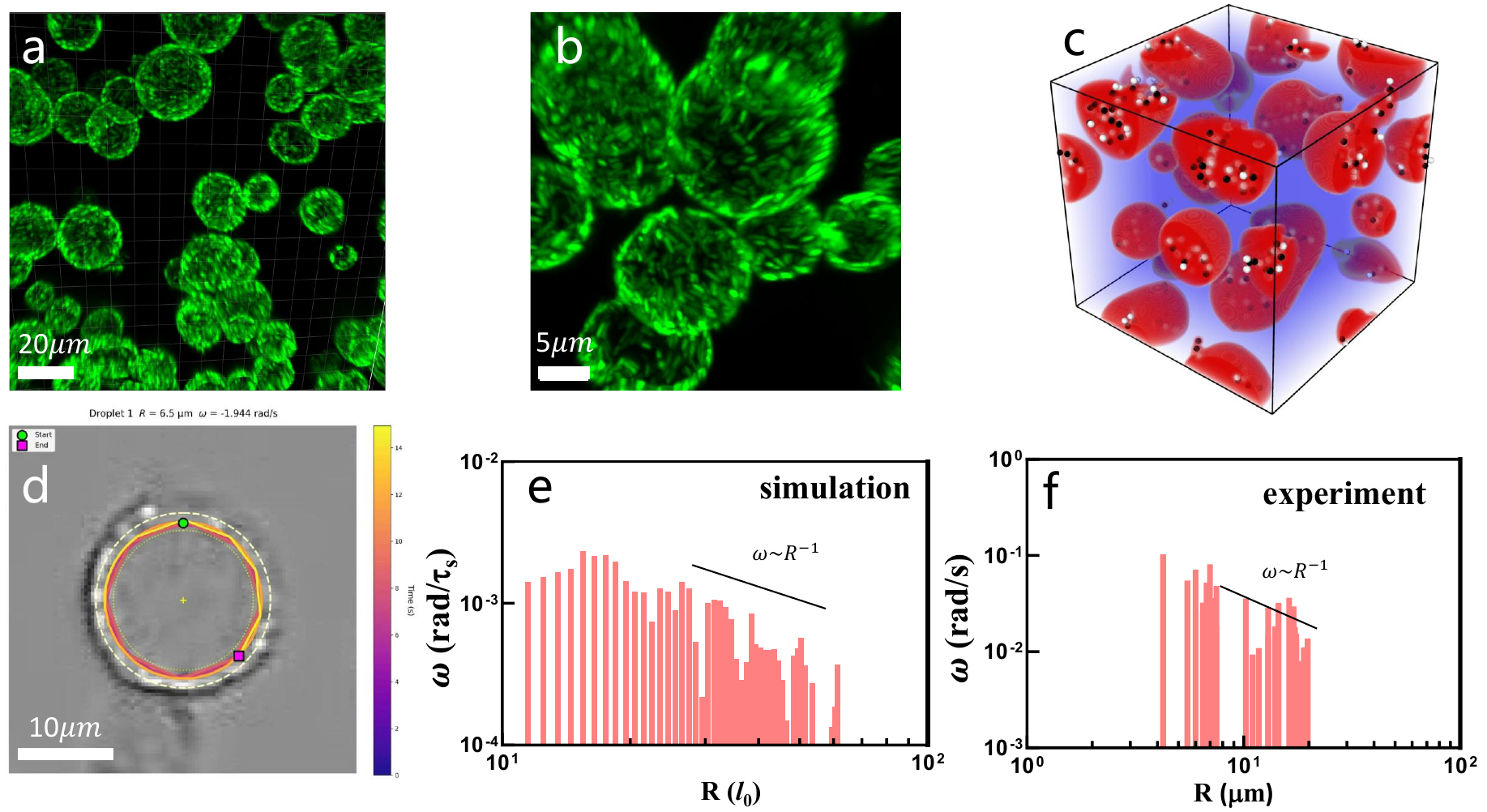}
\caption{Interfacial accumulation and kinematic scaling of active droplets ($\phi_{\mathrm{bac}}=0.01$ and $\phi_{\mathrm{dex}}=0.05$). (a, b) Confocal microscopy images showing the accumulation of GFP-labeled \textit{P. aeruginosa} at the interface of DEX-rich droplets (scale bars: $20~\mu\text{m}$ and $5~\mu\text{m}$, respectively). (c) A 3D snapshot from particle-field hybrid simulations illustrating the partitioning of active bacteria (black and white dots for heads and tails) onto the surface of phase-separated droplets (red). 
The simulation time is $t=5000\tau_s$. (d) Time evolution of a representative DEX-rich droplet with radius $R=6.5~\mu\mathrm{m}$ in experiments. To quantify bacteria-induced rotation, we tracked a surface-functionalized silica tracer particle ($1~\mu$m diameter) pinned at the DEX-rich droplet interface. The tracer trajectory provides a direct measure of the angular velocity $\omega$. (e,f) Log--log plots of $\omega$ versus $R$ obtained from (e) simulations and (f) experiments; the data are consistent with an approximate $R^{-1}$ dependence over the accessible size range.}
\label{fig:figure1}
\end{figure*}

\begin{figure*}[t!]
\centering \includegraphics[width=15cm]{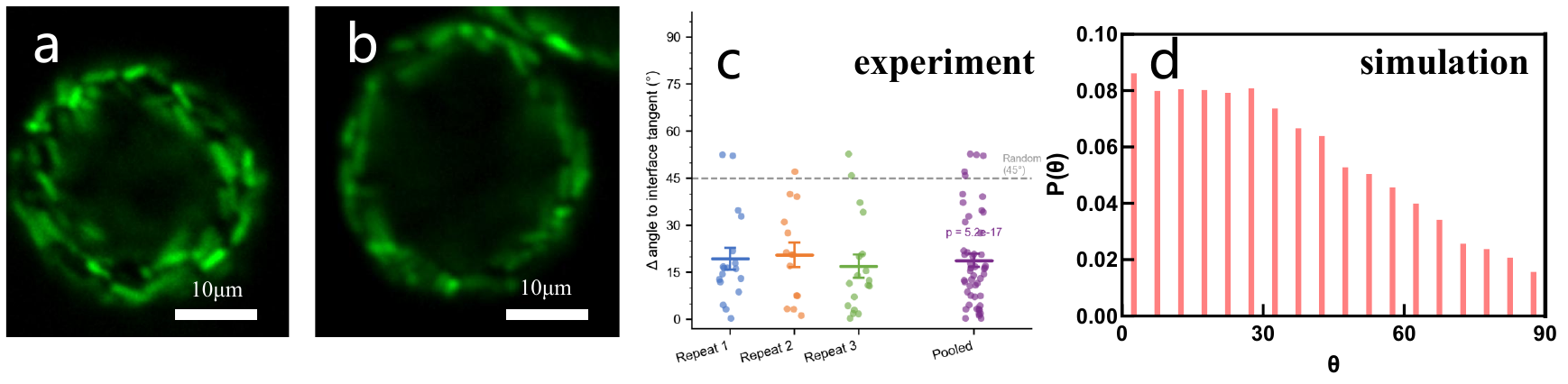}
\caption{Tangential orientational alignment of bacteria at the fluid interface. (a,b) Fluorescence images showing the tangential alignment of individual bacteria along the droplet boundary (scale bars: $10~\mu\text{m}$). (c) Scatter plot of the unsigned bacterial body-orientation angle relative to the interface tangent for three independent experimental repeats. The pooled mean angle of $\sim 20^\circ$ indicates strong preferential tangential alignment, compared with the $45^\circ$ expected for random in-plane orientations ($p_{\rm stat}=5.2\times10^{-17}$). Because this observable is unsigned, it does not by itself establish a signed polar circulation. (d) Probability distribution $P(\theta)$ in simulations, likewise showing preferential tangential alignment.}
\label{fig:figure2}
\end{figure*}

\begin{figure*}[t!]
\centering \includegraphics[width=15cm]{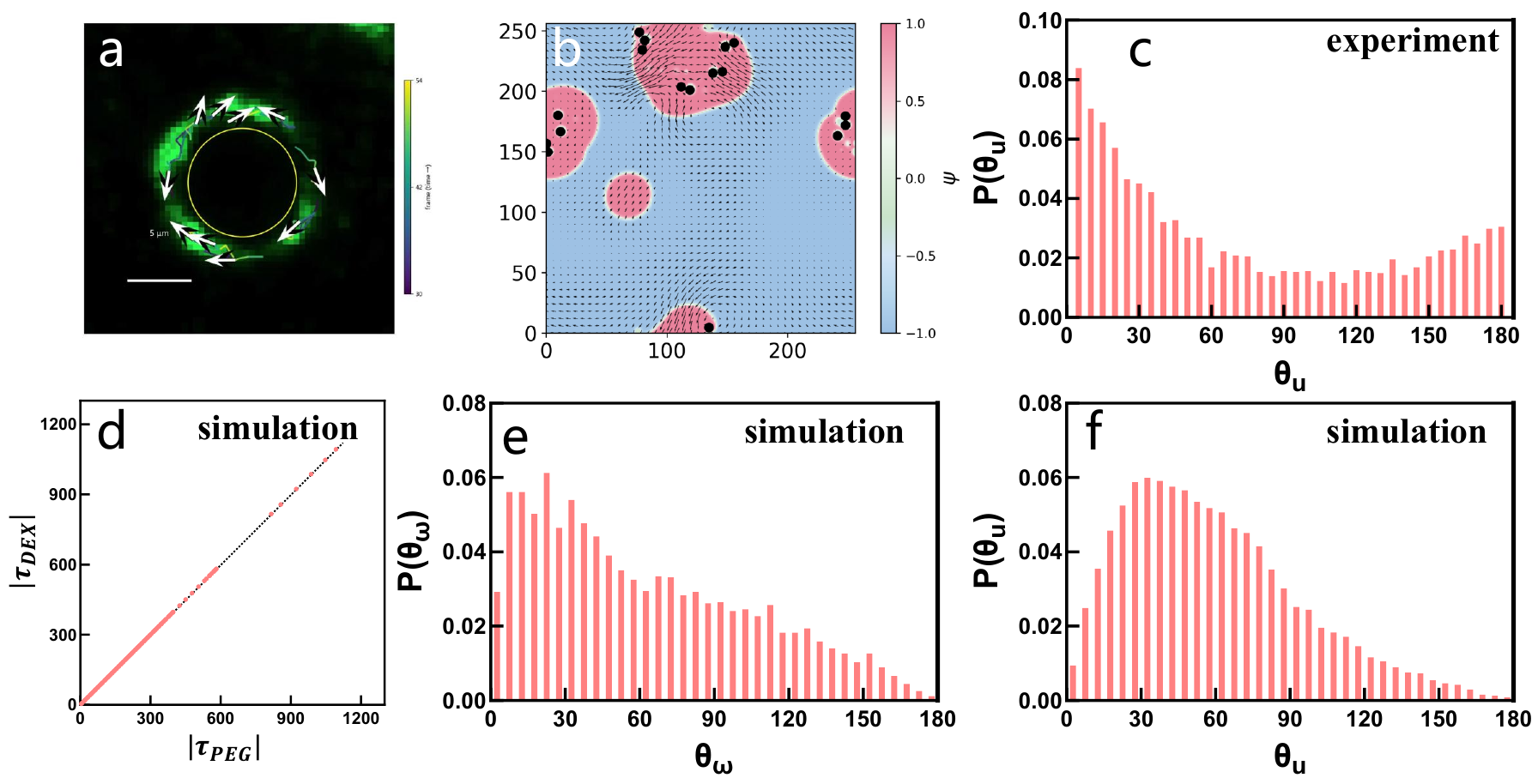}
\caption{Active displacement fields and phase-resolved torque generation by bacteria at the interface of DEX-rich droplets. (a) Experimental bacterial displacement field visualized by white vectors, showing that bacteria predominantly move along the droplet wall to form a coherent tangential flow (scale bar: $5~\mu\text{m}$). (b) Corresponding velocity field from hydrodynamic simulations. (c) Quantitative analysis of the experimental bacterial displacement field, showing strong alignment with the local positive azimuthal direction $\hat{\bm t}_i$. (d) Correlation between the magnitudes of the active torques exerted on the DEX phase ($|\tau_{\text{DEX}}|$) and the PEG phase ($|\tau_{\text{PEG}}|$). The near one-to-one relation between the torque magnitudes is consistent with the partitioning of the torque-free swimmer dipoles between the two phases. The opposite signs of the phase-resolved torques follow from the force-dipole construction, which conserves the total active torque. (e) Probability distribution of the angle $\theta_\omega$ between the droplet angular velocity and $\boldsymbol\tau_{\text{DEX}}$, supporting a mechanical connection between phase-resolved torque and rotation. (f) Probability distribution of the angle $\theta_u$ between bacterial velocity and  local positive azimuthal direction in simulations, showing preferential tangential swimming.}
\label{fig:figure3}
\end{figure*}

\begin{figure*}[t!]
\centering \includegraphics[width=15cm]{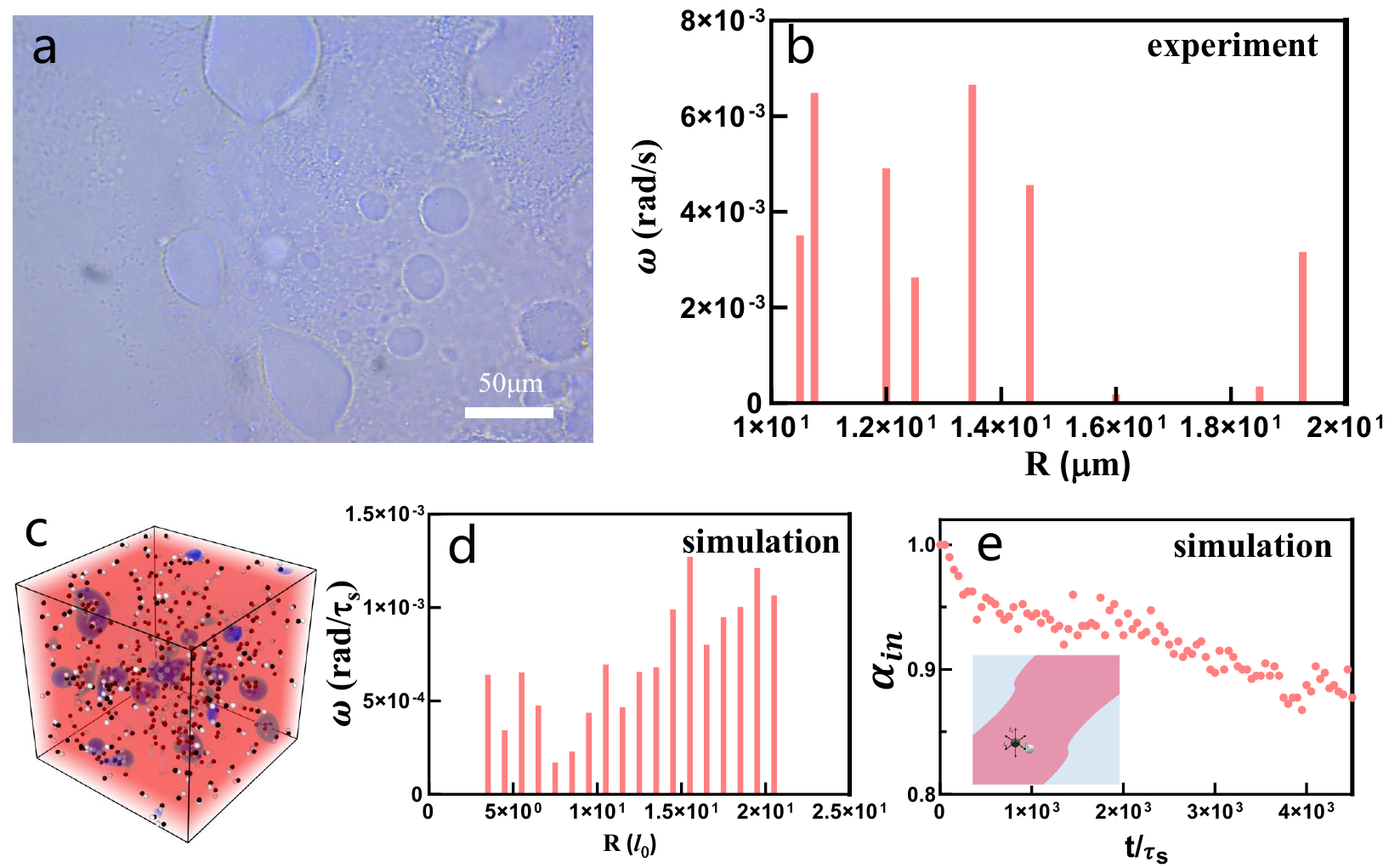}
\caption{Suppressed rotation of PEG-rich droplets. (a,b) Experimental characterization of a DEX--PEG mixture with PEG as the minority phase ($\phi_{\rm peg}=0.05$) in the presence of bacteria ($\phi_{\rm bac}=0.01$). (a) Representative bright-field image showing PEG-rich droplets dispersed in a continuous DEX-rich phase. (b) Measured experimental angular velocity $\omega$ as a function of droplet radius $R$, showing nearly negligible rotational speeds (of order $10^{-3}$ rad\,s$^{-1}$) compared with those of DEX-rich droplets. (c) Simulation snapshot showing that most bacteria remain dispersed in the bulk DEX phase. (d) Simulated angular velocity $\omega$ as a function of droplet radius $R$, demonstrating a strong suppression of rotation relative to DEX-rich droplets. (e) Time evolution of the fraction of bacteria in the DEX phase~$\alpha_\tn{in}$. A substantial fraction of the bacterial population remains distributed in the DEX bulk, consistent with the experimental observation in (a). The inset shows the chord length-based method for calculating~$\alpha_\tn{in}$ (see \cite{Supple} for details).
}
\label{fig:figure4}
\end{figure*}

\subsection{Interfacial localization and droplet rotation}

We first characterized the spatial distribution of \textit{P. aeruginosa} in a phase-separated DEX--PEG ATPS ($\phi_{\mathrm{bac}}=0.01$ and $\phi_{\mathrm{dex}}=0.05$). Confocal microscopy reveals pronounced accumulation of GFP-labeled bacteria at the interfaces of dispersed DEX-rich droplets (Figs.~\ref{fig:figure1}(a) and \ref{fig:figure1}(b)). The three-dimensional particle--field simulations qualitatively reproduce this interfacial localization (Fig.~\ref{fig:figure1}(c)), establishing the structural basis of the model. We note, however, that the instantaneous simulated bacterial coverage is visibly more heterogeneous than in the experiments. Because only a modest number of bacteria occupy an individual simulated droplet interface, part of this patchiness may simply reflect finite-population fluctuations. Persistent gathering, if present after time averaging or at larger interfacial populations, could instead arise from the coupling of bacteria to the composition field, hydrodynamic interactions between active swimmers, or feedback between a locally generated interfacial flow and bacterial advection. Since direct bacterial interactions in the model are repulsive, any persistent clustering would be emergent rather than imposed by an explicit attractive pair potential. The current snapshot alone does not distinguish these possibilities, and we therefore regard the simulation as a qualitative representation of interfacial localization rather than a quantitative reproduction of the experimental surface-density statistics.

Remarkably, this interfacial bacterial layer drives the sustained, autonomous rotation of the droplets ({\bf Movie~S1}). To experimentally quantify the macroscopic angular velocity $\omega$, we tracked the individual tracer particle pinned at the fluid--fluid interface. To achieve this, we specifically optimized the wetting properties of these tracer particles to ensure their stable interfacial pinning. Among the tested silica beads with varying degrees of surface amination (Fig.~S2 in \cite{Supple}), those with 40\% amination exhibited the most stable attachment at the DEX--PEG interface, serving as robust kinematic tracers (Fig.~\ref{fig:figure1}(d)). 

To quantify the rotational dynamics, we analyzed the relationship between $\omega$ and the droplet radius $R$. As shown in Figs.~\ref{fig:figure1}(e) and \ref{fig:figure1}(f), both experiments and simulations show a decrease of $\omega$ with increasing $R$ that is approximately consistent with $\omega\propto R^{-1}$ over the accessible range. Because the experimental and numerical size ranges are limited, we treat this as an approximate trend rather than a precisely established scaling law. In our experiments, the distance between the wall and the droplets used for rotational analysis consistently exceeds 50~$\mu$m (Fig.~S3 in \cite{Supple}), which is substantially larger than the characteristic droplet radius ($R \sim 10~\mu$m). The large droplet--wall separation strongly disfavors wall-mediated rolling as the dominant mechanism of rotation. Furthermore, a control simulation without bacterial motility shows that the droplets remain static and are visibly larger than in the active cases (Fig.~S4 in \cite{Supple}), reflecting the consequences of rotation for droplet-scale coarsening as discussed in our companion work~\cite{yang2026a}. Here, we focus specifically on the physical mechanism underlying this bacteria-driven rotation. Together, the large bulk separation and the activity-free control rule out the possibility that the rotation is induced by nearby solid boundaries or biological factors. Instead, these results support the interpretation that the rotation is a physical consequence driven by interfacial active propulsion.

\subsection{Tangential alignment and global circulation}

We examine the orientation of bacterial bodies along the interface. Figures~\ref{fig:figure2}(a) and \ref{fig:figure2}(b) visualize individual \textit{P. aeruginosa} near the boundary of DEX-rich droplets. Both experiment and simulation show a strong preference for alignment with the local tangent (Figs.~\ref{fig:figure2}(c) and \ref{fig:figure2}(d) for experiments and simulations). The body-orientation observable establishes tangential alignment, indicating the preferred configuration of bacteria parallel to the boundary. However, Fig.~\ref{fig:figure2}(c) does not by itself establish a macroscopic polar state. Interfacial wetting can stabilize a preferred swimmer configuration relative to the interface, while the additional selection of clockwise (CW) versus counter-clockwise (CCW) motion is encoded in the signed circulation $P_{\rm circ}$. 

To distinguish circulation from mere tangential alignment, we introduce a signed circulation order parameter. For an interfacial bacterium $i$, let $\hat{\bm n}_i$ be the outward radial unit vector and $\hat{\bm u}_i$ its unit velocity. For a specified unit rotation axis $\hat{\bm e}$, we define the local positive azimuthal direction $\hat{\bm t}_i=(\hat{\bm e}\times\hat{\bm n}_i)/|\hat{\bm e}\times\hat{\bm n}_i|$ and the corresponding scalar signed circulation
\begin{equation}
P_{\rm circ}^i=\hat{\bm u}_i\cdot\hat{\bm t}_i=\tn{cos}(\theta_u^i)
\label{eq:Pcirc_scalar}
\end{equation}
Here $P_{\rm circ}=\frac{1}{N_s}\sum_{i=1}^{N_s}P_{\rm circ}^i$ measures the degree of coherent azimuthal motion, whereas its sign distinguishes the two senses of circulation relative to the chosen axis. Importantly, $P_{\rm circ}$ can be close to zero even when individual bacteria are strongly aligned with the local tangent, if clockwise and counter-clockwise contributions cancel. Thus, the unsigned body-orientation shown in Fig.~\ref{fig:figure2} measures tangential (axial) alignment but does not by itself establish a nonzero signed circulation.

We analyzed bacterial trajectories at the interface and calculated the statistical distribution of $P_{\rm circ}^i$. The experimentally measured displacement field (Fig.~\ref{fig:figure3}(a)) is strongly biased along the tangential boundary rather than being isotropic. Crucially, the statistics in Fig.~\ref{fig:figure3}(c) reveal that the distribution is predominantly skewed toward acute angles ($\theta_u < 90^\circ$). This preponderance of positive cosine values directly yields a nonzero, macroscopic signed circulation ($P_{\rm circ} = \langle \cos \theta_u \rangle = \int \cos(\theta_u) P(\theta_u) d\theta_u \approx 0.4$). This observation demonstrates that the bacterial motion is not merely tangentially aligned with the interface but exhibits a net directional bias, consistent with coherent collective circulation. Such circulating motion provides a mechanical mechanism for driving droplet rotation through interfacial momentum transfer. Our simulations (Fig.~\ref{fig:figure3}(b)) qualitatively reproduce the rotational flow. The distribution of $\theta_u$ between bacterial velocity and the local positive azimuthal direction similarly demonstrates tangential swimming in simulations (Fig.~\ref{fig:figure3}(f)), qualitatively consistent with the experimental displacement statistics in Fig.~\ref{fig:figure3}(c). Overall, these statistics indicate that despite local orientational fluctuations, the bacterial collective maintains a net macroscopic circulation that drives the droplet to rotate.

\subsection{Phase-resolved torque transfer across interfaces}

To resolve how a globally force- and torque-free swimmer can nevertheless drive one fluid phase, we consider the active torque exerted separately on each phase. Schematically, for phase $\alpha\in\{\mathrm{DEX},\mathrm{PEG}\}$,
\begin{equation}
\boldsymbol\tau_{\alpha}=\int_{\Omega_\alpha}(\bm r-\bm R_c)\times\bm f_{\rm act}(\bm r)\,d\bm r,
\label{eq:phase_torque}
\end{equation}
where $\bm R_c$ is the droplet center and the integration is restricted to the corresponding phase. For a swimmer fully immersed in a homogeneous phase, the collinear propulsion force and hydrodynamic counterforce form a force dipole whose net active torque vanishes. For an interfacially trapped swimmer, by contrast, the two members of the dipole are preferentially transmitted to different phases. Each phase can then experience a finite torque even though the combined system remains torque balanced, with the phase-resolved relation $\boldsymbol\tau_{\rm DEX}+\boldsymbol\tau_{\rm PEG}=0$. Figure~\ref{fig:figure3}(d) supports this physical picture by showing a near one-to-one correspondence between the magnitudes $|\tau_{\rm DEX}|$ and $|\tau_{\rm PEG}|$. Within the force-dipole representation, the total imposed active torque vanishes, so that the phase-resolved active torques are equal and opposite when the two phases
account for the entire active forcing. The numerical results are consistent with this torque balance.
Crucially, the probability distribution of the angle $\theta_\omega$ between the droplet angular velocity~$\omega$ and $\boldsymbol\tau_{\rm DEX}$ (Fig.~\ref{fig:figure3}(e)) is biased toward small angles, supporting the interpretation that the phase-resolved interfacial torque drives the observed rotation. 

Taken together, our results collectively support the following mechanical picture. Interfacial wetting localizes swimmers at the liquid--liquid interface and biases their motion tangentially. The instantaneous torque is then set by the orientational organization of the interfacial population. Because the active force dipole is partitioned across the interface due to interfacial localization of bacteria, the DEX and PEG phases can receive opposite torques even though the total active torque of the combined system remains balanced. The liquid interface is therefore not a source of net angular momentum; rather, it is the mechanical element that spatially partitions the internally generated force dipole and permits phase-resolved torque transfer. This highlights the essential roles of interfacial localization and phase-resolved torque partitioning for driving droplet rotation.

We remark that it is instructive to contrast droplet system with conventional solid boundaries. If these bacterial flocs were swimming along a rigid solid wall, the macroscopic boundary would naturally remain stationary. In stark contrast, the boundary of our system is a highly mobile liquid--liquid interface. It is precisely this fundamental fluidity that allows the microscopic momentum generated by the bacterial propulsion to be dynamically coupled to the boundary, thereby driving the macroscopic rotation of the entire droplet.

\subsection{Theoretical analysis of $\omega$-dependence on $R$}

To rationalize the observed inverse-radius scaling ($\omega \propto R^{-1}$), we develop a minimal
phenomenological model based on hydrodynamic slip and the partitioning of active force dipoles between the two liquid phases. Swimming bacteria act as force-free active dipoles.
When these dipoles straddle the DEX--PEG interface, their forcing can be partitioned asymmetrically between the two phases, generating a finite torque on each phase while preserving the overall torque balance. We characterize the efficiency with which this interfacial forcing drives droplet rotation by a dimensionless parameter $0<\alpha<1$.

For a droplet rotating with signed angular velocity $\omega$, the effective azimuthal force exerted by bacterium $i$ is $f_i=\xi(\alpha v_{\rm bac}P_{\rm circ}^{i}-\omega R)$,
where $\xi$ is an effective hydrodynamic friction coefficient, $v_{\rm bac}$ is the characteristic
interfacial bacterial swimming speed, and $P_{\rm circ}^{i}$ is the signed azimuthal
alignment defined in Eq.~(\ref{eq:Pcirc_scalar}). The signs of $P_{\rm circ}^{i}$ and $\omega$
are defined relative to the same azimuthal reference direction.
The first term represents the circulation-dependent driving force, whereas the second accounts for
the resistance associated with droplet rotation. Averaging over interfacial bacteria gives
$\langle f_i\rangle =\xi(\alpha v_{\rm bac}p-\omega R)$,
where $p=P_{\rm circ}=\langle P_{\rm circ}^{i}\rangle$.

For an interfacial bacterial coverage fraction $\phi$, the total number of bacteria on the
droplet surface is $N=4\pi R^2\phi/A_b$, where $A_b$ is the effective interfacial area
occupied by a bacterium. The resulting collective bacterial torque is
\begin{equation}
\tau_{\rm act} = NR\langle f_i\rangle=\frac{4\pi R^3\phi\xi}{A_b}
(\alpha v_{\rm bac}p-\omega R).
\end{equation}
Balancing this torque against the effective viscous resistance,
$\tau_{\rm visc}=8\pi\eta_{\rm eff}R^3\omega$, where $\eta_{\rm eff}$ is the effective viscosity,
yields
\begin{equation}
\omega = \frac{\alpha v_{\rm bac}p}{R+R_c},
\label{eq:omega_scaling}
\end{equation}
where
\begin{equation}
R_c=\frac{2\eta_{\rm eff}A_b}{\phi\xi}
\end{equation}
is the characteristic crossover length.

To explicitly evaluate $\xi$, we apply the local hydrodynamic drag relation $\xi = c_d \eta_{\rm eff} l_\tn{bac}$ for a bacterium of length $l_\tn{bac}$, where $c_d \sim \mathcal{O}(1)$ is a dimensionless geometric drag coefficient. Approximating the bacterial body as an ellipsoid, its effective projected area on the two-phase interface is $A_b = \frac{\pi}{4} l_\tn{bac} w$, where $w$ is the characteristic bacterial width. Substituting these geometric relations, the macroscopic effective viscosity $\eta_{\rm eff}$ and the bacterial length $l_\tn{bac}$ precisely cancel out. The critical crossover radius therefore reduces to an intrinsic length scale:
\begin{equation}
R_c=\frac{\pi w}{2c_d\phi}.
\end{equation}
This crossover length $R_c$ characterizes the competition between interfacial bacterial
forcing and viscous resistance. In the limit $p\to 0$, the collective driving torque
vanishes and hence $\omega\to 0$. Similarly, decreasing the interfacial coverage
$\phi$ increases $R_c$ and suppresses droplet rotation.

Using the typical bacterial width $w\sim 0.5\,\mu\text{m}$, the interfacial
coverage $\phi\sim 1$, and $c_d\sim\mathcal{O}(1)$, we estimate
$R_c\sim 1\mu\text{m}$. The experimentally investigated droplets
($R\sim 10$--$20\,\mu\text{m}$) are therefore
expected to lie predominantly in the large-radius regime, $R\gg R_c$.
In this regime, the model predicts $\omega\simeq\alpha v_{\rm bac}p/R$,
consistent with the observed approximate scaling $\omega\propto R^{-1}$, provided
that $p$ and the other effective parameters do not vary strongly with droplet size. It is worth noting that while a similar scaling $\omega \propto R^{-1}$ was reported by Dewangan \textit{et al.}~\cite{dewangan2019rotating}, their rotation stems from a distinct wall-induced mechanism rather than interfacial bacterial partitioning in our work.

We note that while our measured interfacial bacterial speed is $v_{\rm bac} \sim 5-10 \, \mu\text{m/s}$, the actual tangential velocity of a typical rotating droplet ($R = 10 \, \mu\text{m}$, $\omega \sim 0.1 \, \text{rad/s}$) is only $\omega R \sim 1 \, \mu\text{m/s}$. This magnitude separation reflects the nearly balanced dipole nature of the swimmers: only a fraction of their propulsion successfully converts into macroscopic rotation. Using the representative experimental values
of $\omega$, $R$, $v_{\rm bac}$, and $p\simeq 0.4$, together with the estimated
crossover length $R_c$, we obtain $\alpha=\omega(R+R_c)/(v_{\rm bac}p)
\sim 0.3$--$0.6$. This estimate indicates that only a fraction of the bacterial swimming speed contributes to the effective azimuthal driving of the droplet. The parameter $\alpha$ should be
regarded as an effective phenomenological coefficient rather than a directly measured
microscopic quantity. Although a first-principle determination of $\alpha$ is beyond the scope of this study, the minimal model provides a physical rationale for the observed approximate
$\omega\propto R^{-1}$ dependence. For a circulation state with $p>0$, the model predicts
$\omega R=\alpha v_{\rm bac}pR/(R+R_c) <\alpha v_{\rm bac}p$.
Thus, the droplet's tangential velocity remains below the circulation-weighted effective
bacterial swimming speed. This velocity mismatch reflects the energy balance of the system: the bacteria must move faster than the droplet surface to continuously pump energy into the system, balancing the viscous energy dissipation of the rotation.

\subsection{Suppression of rotation in bulk-dominated control systems}

Finally, we test the importance of interfacial localization using an inverted configuration in which PEG is the minority phase ($\phi_{\rm peg}=0.05$). As shown in Figs.~\ref{fig:figure4}(a) and \ref{fig:figure4}(b), PEG-rich droplets rotate much more slowly than DEX-rich droplets, with measured angular velocities smaller by roughly an order of magnitude. Because bacteria preferentially occupy the DEX-rich environment, the continuous DEX-rich phase provides a large bulk reservoir and only a reduced fraction of bacteria accumulates at the PEG-droplet interface ({\bf Movie~S2}). The FPD simulations successfully reproduce this redistribution and the concomitant suppression of rotation (Figs.~\ref{fig:figure4}(c)--(e)). This observation is consistent with
Eq.~(\ref{eq:omega_scaling}), which predicts that reducing the interfacial coverage
($\phi\to0$) suppresses droplet rotation ($\omega\to0$).

\section{Conclusion}

To summarize, we combine bacterial experiments, coarse-grained hydrodynamic simulations, and theoretical analysis to uncover a mechanism by which interfacial bacteria can drive autonomous rotation of droplets in a three-dimensional aqueous two-phase system. The measurements establish strong interfacial accumulation and tangential swimmer organization, while the simulations show how an interfacially straddling force dipole can generate finite, oppositely balanced torques on the DEX- and PEG-rich phases. This phase-resolved torque transfer provides a direct mechanical route from microscopic propulsion to droplet-scale rotation without requiring a bulk active-nematic state. Over the accessible size range, the decrease of angular velocity with droplet size is approximately consistent with an $R^{-1}$ dependence. By formulating a theoretical model, we rationalize this scaling law and demonstrate that  droplet rotation is controlled by the interfacial bacterial coverage and the emergence of collective circular flow. Crucially, our orientational statistics provide evidence that droplet rotation
is associated with a coherent circulating state of interfacial bacteria, rather
than merely random tangential motion. More generally, our results identify fluid--fluid interfaces as mechanical partitioners of active force dipoles and provide a practical route to regulate
droplet dynamics via interfacial active matter.

While our work establishes the emergence of macroscopic circulation driven by interfacial bacteria, the microscopic origin of the selection of a particular rotational direction presents an open avenue for future investigation. Possible mechanisms likely involve a complex interplay between collective many-body alignment and hydrodynamic feedback that amplifies initially small directional fluctuations. Disentangling these mechanisms remains a compelling question for future experimental investigations.

\begin{acknowledgments}
We thank Huihui Liu for performing part of the experiments and Tianyi Zhang for contributing to the partial implementation of the code.  J.Y. is supported by the National Natural Science Foundation of China (Grant No. 22503076) and the Guangdong Provincial Pearl River Talents Program (Grant No. 2024GA11X007). C.W. is supported by the Science and Technology Development Fund, Macao SAR (FDCT, Nos. 0010/2026/AIJ, 0039/2025/RIB1, 0024/2023/AFJ, and 0002/2025/NRP), the Guangdong Provincial Committee for Basic and Applied Basic Research (EF2023-00209-ICMS), as well as the Zhuhai UM Science \& Technology Research Institute (CP-102-2024). H.T. acknowledges support from the Grant-in-Aid for Specially Promoted Research (KAKENHI Grant No. JP20H05619) and Scientific Research (A) (KAKENHI Grant No. JP26H02008) from the Japan Society for the Promotion of Science (JSPS). J.Y. and D.Y. thank the HPC AI Intelligent Computing Platform at HKUST(GZ) for GPU resources. 
\end{acknowledgments}

\bibliography{journals,bac,vps}

@string{acsnano = {ACS Nano}}

@string{advmat = {Adv. Mater.}}

@string{jacs = {J. Am. Chem. Soc.}}

@string{natcomm = {Nat. Commun.}}

@string{natphys = {Nat. Phys.}}

@string{pnas = {Proc. Natl. Acad. Sci. U.S.A.}}

@string{pre = {Phys. Rev. E}}

@string{prl = {Phys. Rev. Lett.}}

@string{softmat = {Soft Matter}}

@article{schwarz2012phase,
  title={Phase separation and rotor self-assembly in active particle suspensions},
  author={Schwarz-Linek, J. and Valeriani, C. and Cacciuto, A. and Cates, M. E. and Marenduzzo, D. and Morozov, A. N. and Poon, W. C. K.},
  journal=pnas,
  volume={109},
  number={11},
  pages={4052--4057},
  year={2012},
  publisher={National Academy of Sciences}
}

@article{ramos2020bacteria,
  title={Bacteria driving droplets},
  author={Ramos, Gabriel and Cordero, Mar{\'\i}a Luisa and Soto, Rodrigo},
  journal=softmat,
  volume={16},
  number={5},
  pages={1359--1365},
  year={2020},
  publisher={Royal Society of Chemistry}
}

@article{rajabi2021directional,
  title={Directional self-locomotion of active droplets enabled by nematic environment},
  author={Rajabi, Mojtaba and Baza, Hend and Turiv, Taras and Lavrentovich, Oleg D},
  journal=natphys,
  volume={17},
  number={2},
  pages={260--266},
  year={2021},
  publisher={Nature Publishing Group UK London}
}

@article{furukawa2014activity,
  title={Activity-induced clustering in model dumbbell swimmers: {T}he role of hydrodynamic interactions},
  author={Furukawa, Akira and Marenduzzo, Davide and Cates, Michael E},
  journal=pre,
  volume={90},
  number={2},
  pages={022303},
  year={2014},
  publisher={APS}
}

@article{yuan2024colloid,
  title = {Colloid phase separation dynamics driven by chiral turbulent flows},
  author = {Yuan, Jiaxing and Tanaka, Hajime},
  journal = {Phys. Rev. Res.},
  volume = {6},
  issue = {2},
  pages = {023186},
  numpages = {11},
  year = {2024},
  month = {May}
}

@article{cheon2024partitioning,
  title = {Motility Modulates the Partitioning of Bacteria in Aqueous Two-Phase Systems},
  author = {Cheon, Jiyong and Choi, Kyu Hwan and Modica, Kevin J. and Mitchell, Robert J. and Takatori, Sho C. and Jeong, Joonwoo},
  journal = {Phys. Rev. Lett.},
  volume = {135},
  issue = {12},
  pages = {128401},
  numpages = {6},
  year = {2025},
  month = {Sep},
  publisher = {American Physical Society}
}

@article{grober2023unconventional,
  title={Unconventional colloidal aggregation in chiral bacterial baths},
  author={Grober, Daniel and Palaia, Ivan and U{\c{c}}ar, Mehmet Can and Hannezo, Edouard and {\v{S}}ari{\'c}, Andela and Palacci, J{\'e}r{\'e}mie},
  journal={Nat. Phys.},
  volume={19},
  number={11},
  pages={1680--1688},
  year={2023},
  publisher={Nature Publishing Group UK London}
}

@article{drescher2011fluid,
  title={Fluid dynamics and noise in bacterial cell--cell and cell--surface scattering},
  author={Drescher, Knut and Dunkel, J{\"o}rn and Cisneros, Luis H and Ganguly, Sujoy and Goldstein, Raymond E},
  journal=pnas,
  volume={108},
  number={27},
  pages={10940--10945},
  year={2011},
  publisher={National Acad Sciences}
}

@article{wang2019progress,
author = {Wang, Fei and Altschuh, Patrick and Ratke, Lorenz and Zhang, Haodong and Selzer, Michael and Nestler, Britta},
title = {Progress Report on Phase Separation in Polymer Solutions},
journal = advmat,
volume = {31},
number = {26},
pages = {1806733},
year = {2019}
}

@book{onuki2002phase,
  title={Phase transition dynamics},
  author={Onuki, Akira},
  year={2002},
  publisher={Cambridge University Press}
}

@article{iborra2007can,
  title={Can viscoelastic phase separation, macromolecular crowding and colloidal physics explain nuclear organisation?},
  author={Iborra, Francisco J},
  journal={Theor. Biol. Medical Model.},
  volume={4},
  pages={15},
  year={2007},
  publisher={BioMed Central}
}

@article{yuan2025network,
  title={Network-forming phase separation of oppositely charged polyelectrolytes forming coacervates in a solvent},
  author={Yuan, Jiaxing and Tanaka, Hajime},
  journal=natcomm,
  volume={16},
  number={1},
  pages={1517},
  year={2025},
  publisher={Nature Publishing Group UK London}
}

@article{arnold2023active,
  title = {Active Surface Flows Accelerate the Coarsening of Lipid Membrane Domains},
  author = {Arnold, Daniel P. and Gubbala, Aakanksha and Takatori, Sho C.},
  journal = prl,
  volume = {131},
  issue = {12},
  pages = {128402},
  numpages = {6},
  year = {2023},
  month = {Sep},
  publisher = {American Physical Society}
}

@article{Yuan2023Mechanical,
author = {Yuan, Jiaxing and Tateno, Michio and Tanaka, Hajime},
title = {Mechanical Slowing Down of Network-Forming Phase Separation of Polymer Solutions},
journal = acsnano,
volume = {17},
number = {18},
pages = {18025-18036},
year = {2023}
}

@article{shaebani2020computational,
  title={Computational models for active matter},
  author={Shaebani, M Reza and Wysocki, Adam and Winkler, Roland G and Gompper, Gerhard and Rieger, Heiko},
  journal={Nat. Rev. Phys.},
  volume={2},
  number={4},
  pages={181--199},
  year={2020},
  publisher={Nature Publishing Group UK London}
}

@article{Singh2019,
  title = {Hydrodynamically Interrupted Droplet Growth in Scalar Active Matter},
  author = {Singh, Rajesh and Cates, M. E.},
  journal = prl,
  volume = {123},
  issue = {14},
  pages = {148005},
  numpages = {6},
  year = {2019},
  month = {Oct},
  publisher = {American Physical Society}
}

@misc{Supple,
    howpublished = {See Supplemental Material at
  http://link.aps.org/supplemental/xxx on the details of experiments and simulations and the additional characterization.}
    }

@article{yang2026a,
title = {Active motility and wetting cooperatively regulate liquid--liquid phase separation},
journal = {Submitted to Physical Review Letters},
author = {Yang, Dixi and Wang, Anheng and Wang, Chunming and  Tanaka, Hajime and Yuan, Jiaxing}
}

@article{marchetti2013hydrodynamics,
  author = {Marchetti, M. C. and Joanny, J. F. and Ramaswamy, S. and
            Liverpool, T. B. and Prost, J. and Rao, Madan and Simha, R. Aditi},
  title = {Hydrodynamics of soft active matter},
  journal = {Rev. Mod. Phys.},
  volume = {85},
  number = {3},
  pages = {1143--1189},
  year = {2013}
}

@article{bechinger2016active,
  author = {Bechinger, Clemens and Di Leonardo, Roberto and L{\"o}wen, Hartmut and
            Reichhardt, Charles and Volpe, Giorgio and Volpe, Giovanni},
  title = {Active particles in complex and crowded environments},
  journal = {Rev. Mod. Phys.},
  volume = {88},
  number = {4},
  pages = {045006},
  year = {2016}
}

@article{michelin2023self,
  author = {Michelin, S{\'e}bastien},
  title = {Self-propulsion of chemically active droplets},
  journal = {Annu. Rev. Fluid Mech.},
  volume = {55},
  pages = {77--101},
  year = {2023}
}

@article{gao2017self,
  author = {Gao, Tong and Li, Zhaorui},
  title = {Self-driven droplet powered by active nematics},
  journal = {Phys. Rev. Lett.},
  volume = {119},
  number = {10},
  pages = {108002},
  year = {2017}
}

@article{morozov2019orientational,
  author = {Morozov, Matvey and Michelin, S{\'e}bastien},
  title = {Orientational instability and spontaneous rotation of active nematic droplets},
  journal = {Soft Matter},
  volume = {15},
  number = {39},
  pages = {7814--7822},
  year = {2019}
}

@article{sokolov2010swimming,
  author = {Sokolov, Andrey and Apodaca, Mario M. and Grzybowski, Bartosz A. and
            Aranson, Igor S.},
  title = {Swimming bacteria power microscopic gears},
  journal = {Proc. Natl. Acad. Sci. U.S.A.},
  volume = {107},
  number = {3},
  pages = {969--974},
  year = {2010}
}

@article{dileonardo2010bacterial,
  author = {Di Leonardo, R. and Angelani, L. and Dell'Arciprete, D. and
            Ruocco, G. and Iebba, V. and Schippa, S. and Conte, M. P. and
            Mecarini, F. and De Angelis, F. and Di Fabrizio, E.},
  title = {Bacterial ratchet motors},
  journal = {Proc. Natl. Acad. Sci. U.S.A.},
  volume = {107},
  number = {21},
  pages = {9541--9545},
  year = {2010}
}

@article{vincenti2019magnetotactic,
  author = {Vincenti, Benoit and Ramos, Gabriel and Cordero, Maria Luisa and
            Douarche, Carine and Soto, Rodrigo and Cl{\'e}ment, Eric},
  title = {Magnetotactic bacteria in a droplet self-assemble into a rotary motor},
  journal = {Nat. Commun.},
  volume = {10},
  pages = {5082},
  year = {2019}
}

@article{dewangan2019rotating,
  author = {Dewangan, Narendra K. and Conrad, Jacinta C.},
  title = {Rotating oil droplets driven by motile bacteria at interfaces},
  journal = {Soft Matter},
  volume = {15},
  number = {45},
  pages = {9368--9375},
  year = {2019}
}

@article{vaccari2015films,
  author = {Vaccari, Liana and Allan, Daniel B. and Sharifi-Mood, Nima and
            Singh, Aayush R. and Leheny, Robert L. and Stebe, Kathleen J.},
  title = {Films of bacteria at interfaces: three stages of behaviour},
  journal = {Soft Matter},
  volume = {11},
  number = {30},
  pages = {6062--6074},
  year = {2015}
}

@article{vaccari2017films,
  author = {Vaccari, Liana and Molaei, Mehdi and Niepa, Tagbo H. R. and
            Lee, Daeyeon and Leheny, Robert L. and Stebe, Kathleen J.},
  title = {Films of bacteria at interfaces},
  journal = {Adv. Colloid Interface Sci.},
  volume = {247},
  pages = {561--572},
  year = {2017}
}

@article{niepa2017films,
  author = {Niepa, Tagbo H. R. and Vaccari, Liana and Leheny, Robert L. and
            Goulian, Mark and Lee, Daeyeon and Stebe, Kathleen J.},
  title = {Films of bacteria at interfaces ({FBI}): Remodeling of fluid
           interfaces by {{\textit{Pseudomonas aeruginosa}}}},
  journal = {Sci. Rep.},
  volume = {7},
  pages = {17864},
  year = {2017}
}

@article{deng2020motile,
  author = {Deng, Jiayi and Molaei, Mehdi and Chisholm, Nicholas G. and
            Stebe, Kathleen J.},
  title = {Motile bacteria at oil--water interfaces:
           {{\textit{Pseudomonas aeruginosa}}}},
  journal = {Langmuir},
  volume = {36},
  number = {25},
  pages = {6888--6902},
  year = {2020}
}

@article{liu2020oscillating,
  title={Oscillating collective motion of active rotors in confinement},
  author={Liu, Peng and Zhu, Hongwei and Zeng, Ying and Du, Guangle and Ning, Luhui and Wang, Dunyou and Chen, Ke and Lu, Ying and Zheng, Ning and Ye, Fangfu and others},
  journal=pnas,
  volume={117},
  number={22},
  pages={11901--11907},
  year={2020},
  publisher={National Acad Sciences}
}

@article{Yuan2022, 
  title = {Impact of Inverse Squeezing Flow on the Self-Assembly of Oppositely Charged Colloidal Particles under Electric Field},
  author = {Yuan, Jiaxing and Takae, Kyohei and Tanaka, Hajime},
  journal = prl,
  volume = {129},
  issue = {24},
  year = {2022},
  pages = {248001}
}

@STRING{acsnano	= "ACS Nano" }

@STRING{advmat	= "Adv. Mater." }

@STRING{jacs	= "J. Am. Chem. Soc." }

@STRING{natcomm	= "Nat. Commun." }

@STRING{natphys = "Nat. Phys." }

@STRING{nature	= "Nature" }

@STRING{physics	= "Physics" }

@STRING{pnas	= "Proc. Natl. Acad. Sci. U.S.A." }

@STRING{polymer	= "Polymer" }

@STRING{pre	= "Phys. Rev. E" }

@STRING{prl	= "Phys. Rev. Lett." }

@STRING{softmat	= "Soft Matter" }

@article{chen2026entropic,
  title={Entropic Charge Separation as a General Mechanism Arresting Nanoscale Condensate Coarsening},
  author={Chen, Feipeng and Yuan, Jiaxing and Zhang, Yaojun and Tanaka, Hajime and Shum, Ho Cheung},
  journal=prl,
  volume={136},
  number={22},
  pages={228202},
  year={2026},
  publisher={APS}
}

@article{chen2026phase,
  title={Phase separation-engineered cell assembly into functional tissue architecture},
  author={Chen, Chonghao and Gu, Xiangyang and Liao, Zhencheng and Wang, Anheng and Xie, Daping and Dong, Lei and Chen, Guokai and Yuan, Jiaxing and Tanaka, Hajime and Wang, Chunming},
  journal={Cell Biomaterials},
  year={2026},
  publisher={Elsevier}
}

@article{yang2025active,
  title={Active motility and wetting cooperatively regulate liquid-liquid phase separation},
  author={Yang, Dixi and Wang, Anheng and Wang, Chunming and Tanaka, Hajime and Yuan, Jiaxing},
  journal={arXiv preprint arXiv:2511.18077},
  year={2025}
}

@article{Harlow1965,
author = {Harlow,Francis H. and Welch,J. Eddie},
title = {Numerical Calculation of Time-Dependent Viscous Incompressible Flow of Fluid with Free Surface},
journal = {Phys. Fluids},
volume = {8},
number = {12},
pages = {2182--2189},
year = {1965}
}

@article{		  tanaka2022Viscoelastic,
  author	= "Tanaka, Hajime",
  title		= "Viscoelastic Phase Separation in Biological Cells",
  journal = {Commun. Phys.},
  year		= "2022",
  volume = {5},
  pages = {167}
}

@article{Zhang2021,
  title = {Mechanical Frustration of Phase Separation in the Cell Nucleus by Chromatin},
  author = {Zhang, Yaojun and Lee, Daniel S. W. and Meir, Yigal and Brangwynne, Clifford P. and Wingreen, Ned S.},
  journal = prl,
  volume = {126},
  issue = {25},
  pages = {258102},
  numpages = {6},
  year = {2021},
  month = {Jun},
  publisher = {American Physical Society}
}

@article{tanaka2000,
  title = {Simulation Method of Colloidal Suspensions with Hydrodynamic Interactions: {F}luid Particle Dynamics},
  author = {Tanaka, Hajime and Araki, Takeaki},
  journal = prl,
  volume = {85},
  issue = {6},
  pages = {1338--1341},
  numpages = {0},
  year = {2000},
  month = {Aug}
}

@Article{Furukawa2018,
author ="Furukawa, Akira and Tateno, Michio and Tanaka, Hajime",
title  ="Physical Foundation of the Fluid Particle Dynamics Method for Colloid Dynamics Simulation",
journal  = softmat,
year  ="2018",
volume  ="14",
issue  ="19",
pages  ="3738--3747",
publisher  ="The Royal Society of Chemistry"
}

@article{berry2018physical,
  title={Physical Principles of Intracellular Organization via Active and Passive Phase Transitions},
  author={Berry, Joel and Brangwynne, Clifford P and Haataja, Mikko},
  journal={Rep. Prog. Phys.},
  volume={81},
  number={4},
  pages={046601},
  year={2018},
  publisher={IOP Publishing}
}

@article{schmid2022understanding,
author = {Schmid, Friederike},
title = {Understanding and Modeling Polymers: The Challenge of Multiple Scales},
journal = {ACS Polymers Au},
volume = {3},
number = {1},
pages = {28--58},
year = {2023}
}

@article{glotzer1993physical,
  title={Physical Gels and Microphase Separation in Multiblock Copolymers},
  author={Glotzer, Sharon C and Bansil, Rama and Gallagher, Patrick D and Gyure, Mark F and Sciortino, Francesco and Stanley, H Eugene},
  journal={Physica A},
  volume={201},
  number={4},
  pages={482--495},
  year={1993},
  publisher={Elsevier}
}

@article{Ebbens2011,
author = {Ebbens, Stephen and Hodgkinson, Richard and Parnell, Andrew J. and Dunbar, Alan and Martin, Simon J. and Topham, Paul D. and Clarke, Nigel and Howse, Jonathan R.},
title = {In{S}itu Imaging and Height Reconstruction of Phase Separation Processes in Polymer Blends during Spin Coating},
journal = {ACS Nano},
volume = {5},
number = {6},
pages = {5124--5131},
year = {2011}
}

@article{Bok2022,
author = {Bok, Moonjeong and Zhao, Zhi-Jun and Hwang, Soon Hyoung and Ahn, Junseong and Ko, Jiwoo and Jung, Joo-Yun and Lee, Jihye and Jeon, Sohee and Jeong, Jun-Ho},
title = {Functional Asymmetry-Enabled Self-Adhesive Film via Phase Separation of Binary Polymer Mixtures for Soft Bio-Integrated Electronics},
journal = {ACS Nano},
volume = {16},
number = {11},
pages = {18157--18167},
year = {2022}
}

@article{liu2021evolution,
  title={Evolution of Polymer Colloid Structure During Precipitation and Phase Separation},
  author={Liu, Jason X and Bizmark, Navid and Scott, Douglas M and Register, Richard A and Haataja, Mikko P and Datta, Sujit S and Arnold, Craig B and Priestley, Rodney D},
  journal={JACS Au},
  volume={1},
  number={7},
  pages={936--944},
  year={2021},
  publisher={ACS Publications}
}

@article{araki2006wetting,
  title={Wetting-induced depletion interaction between particles in a phase-separating liquid mixture},
  author={Araki, Takeaki and Tanaka, Hajime},
  journal=pre,
  volume={73},
  number={6},
  pages={061506},
  year={2006},
  publisher={APS}
}

\end{document}